\documentclass[a4paper,11pt]{article}
\usepackage{pos}
\usepackage{bm}
\usepackage{multirow}

\usepackage{tikz}
\def\checkmark{\tikz\draw[scale=0.4,fill=black](0,.35) -- (.25,0) -- (1,.7) -- (.25,.15) -- cycle;}
\def\halfcheckmark{\tikz\draw[scale=0.4,fill=black](0,.35) -- (.25,0) -- (1,.7) -- (.25,.15) -- cycle (0.75,0.2) -- (0.77,0.2) -- (0.6,0.7) -- cycle;}
\title{Prospects for measuring quark polarization and spin correlations in $b\bar{b}$ and $c\bar{c}$ samples at the LHC}
\ShortTitle{Prospects for measuring quark polarization and spin correlations in $b\bar{b}$ and $c\bar{c}$ samples}

\author{Yevgeny Kats}
\author*{David Uzan}

\affiliation{Department of Physics, Ben-Gurion University, Beer-Sheva 8410501, Israel}

\emailAdd{katsye@bgu.ac.il}
\emailAdd{daviduz@post.bgu.ac.il}

\abstract{Polarization and spin correlations have been explored very little for quarks other than the top. Utilizing the partial preservation of the quark's spin information in baryons in the jet produced by the quark, we examine possible analysis strategies for ATLAS and CMS to measure the quark polarization and spin correlations in $pp\to q\bar{q}$ processes. We find polarization measurements for the $b$ and $c$ quarks to be feasible, even with the currently available datasets. Spin correlation measurements for $b\bar{b}$ are possible using the CMS Run~2 parked data, while such measurements for $c\bar{c}$ will become possible with higher integrated luminosity. We also provide leading-order QCD predictions for the polarization and spin correlations expected in the $b\bar{b}$ and $c\bar{c}$ samples with the relevant cuts. 
These proposed analyses can provide new information on the polarization transfer from quarks to baryons, measure entanglement and Bell nonlocality in the $b\bar b$ final state, and might even be sensitive to physics beyond the Standard Model.}

\FullConference{%
  42nd International Conference on High Energy Physics (ICHEP2024)\\
  18-24 July 2024\\
  Prague, Czech Republic
}

\begin{document}
\maketitle

\section{Introduction}

Polarization and spin correlation measurements are well-established for top-quark pairs (e.g. Ref.~\cite{CMS:2019nrx}), but have not yet been pursued in detail for pairs of other quarks. In our work~\cite{Kats:2023zxb}, we propose methods for doing such measurements for the heavy quarks $q = b, c$ through their hadronization to $\Lambda_q$ baryons. 
Although the quarks hadronize, the hadronization products can be used to infer the quarks' polarizations and spin correlations. Because the primary spin interaction of a heavy quark occurs via its chromomagnetic dipole moment $\mu_q\propto 1/m_q$ with $m_q\gg\Lambda_{\rm QCD}$, the polarization loss in hadronization is small. Moreover, in the quark model, the $\Lambda_q$ baryon's spin is determined by the spin of the heavy quark $q$ since the light quarks form a spin-0 state. Consequently, the $\Lambda_q$ baryons carry a large fraction of the original quark polarization~\cite{Mannel:1991bs,Ball:1992fw,Falk:1993rf,Galanti:2015pqa}. Measurements of $b$-quark polarization have been performed, although with large statistical uncertainties, in $Z\to b\bar b$ decays at LEP, using the semileptonic decays of the $\Lambda_b$~\cite{ALEPH:1995aqx,OPAL:1998wmk,DELPHI:1999hkl}. We examine Run~2 simulated data from the LHC, with ATLAS and CMS trigger motivated cuts, as well as the expected results from the High Luminosity LHC (HL-LHC).

A $q \bar q$ pair is a system of two spin-$1/2$ particles. It is described by the density matrix 
\begin{equation}
\rho=\frac{I_4+\sum_{i}\left(b^{+}_i\sigma^i\otimes I_2+b^{-}_i I_2\otimes\sigma^i \right)+\sum_{i,j}c_{ij}\sigma^{i}\otimes\sigma^{j}}{4} \,,
\end{equation}
where $I_n$ is the $n\times n$ identity matrix, $b^{\pm}_i$ are the components of the Bloch vectors $\mathbf{b}^{\pm}$ that represent the quark/antiquark polarizations, $c_{ij}$ are the elements of the spin-correlation matrix $\mathbf{c}$, and $\sigma^i$ are the Pauli matrices. In~\cite{Kats:2023zxb} we give leading-order QCD predictions for $b_i^\pm$ and $c_{ij}$ in $b\bar b$ and $c \bar c$ samples.

Since $\Lambda_q$ is the most common $q$-baryon, it is a useful proxy for the quark polarization and spin correlations. 
Polarization and spin-correlation measurements for $b$ quarks can be performed with the $\Lambda_b$ and $\overline\Lambda_b$ decaying semileptonically via $\Lambda_b \to X_c\ell^-\bar{\nu}_\ell$, where $X_c$ denotes a charmed state containing a baryon, usually the $\Lambda_c^+$. 
The angular distributions of the neutrinos from the two semileptonic decays approximately have the shapes
\begin{equation}
\frac{1}{\sigma}\frac{d\sigma}{d\cos\theta^\pm_i} =
    \frac{1}{2}\left(1 + B^\pm_i\cos\theta^\pm_i\right)\,, \quad\frac{1}{\sigma}\frac{d\sigma}{dx_{ij}} =
\frac{1}{2}\left(1 - C_{ij}x_{ij}\right) \ln\left(\frac{1}{|x_{ij}|}\right) ,
\end{equation}
where the angles $\theta^+_i$ ($\theta^-_j$) describe the directions of the antineutrino (neutrino) from the $\Lambda_b$ ($\overline\Lambda_b$) decay with respect to the $i$ ($j$) axis in the $b\bar b$ center-of-mass frame, $x_{ij} = \cos\theta^+_i\cos\theta^-_j$, and
\begin{equation}
B^\pm_i = \alpha r_i f b^\pm_i\,, \quad C_{ij} = \alpha^2 r_i r_j f c_{ij} \,.
\end{equation}
Here, $\alpha \simeq 1$ is the spin analyzing power of the (anti)neutrino in the semileptonic $\Lambda_b$ decay, and $f$ is the sample purity. The factors $r_i$ and $r_j$ are the longitudinal ($r_L$, along the fragmentation axis) or transverse ($r_T$, perpendicular to the fragmentation axis) polarization retention factors~\cite{Falk:1993rf,Galanti:2015pqa}. They describe the fraction of the original quark's longitudinal or transverse polarization that is retained in the baryon and are rough approximations of the spin-dependent fragmentation functions (e.g.~\cite{Metz:2016swz,Chen:1994ar}). Their values are expected to be roughly $r_L, r_T \sim 0.5$~\cite{Galanti:2015pqa}, where the dominant polarization loss arises from $\Lambda_b$ baryon production in $\Sigma_b^{(\ast)}$ decays~\cite{Falk:1993rf,Galanti:2015pqa}. An approximate combination of the LEP measurements~\cite{ALEPH:1995aqx,OPAL:1998wmk,DELPHI:1999hkl} gives $r_L = 0.47 \pm 0.14$ for the $b$ quarks. 
It is also possible to measure $r_L$ for both $b$ and $c$ quarks with the data already available in ATLAS and CMS, in $t\bar t$ samples, using the highly polarized $b$ quarks from the $t\to W^+b$ decays and $c$ quarks from the $W^+\to c\bar s$ decays~\cite{Galanti:2015pqa}. For $c$ quarks, a measurement in $W$+$c$ samples may be possible as well~\cite{Kats:2015zth}.

Measuring $B_i^\pm$ and $C_{ij}$ requires reconstructing the momenta of the quarks, $\Lambda_q$ baryons, and the relevant decay products. This can be done approximately as outlined in Refs.~\cite{Galanti:2015pqa,Dambach:2006ha,LHCb:2015eia,LHCb:2020ist,Ciezarek:2016lqu,Dambach:2009wda}.

For $c$ quarks, we consider the $\Lambda_c^+\to \Lambda(\to p\pi^-)\ell^+\nu_\ell$ decay. The angular distributions and their coefficients are described by the same expressions except that here we use the charged lepton instead of the neutrino as the spin analyzer, which in this case has $\alpha_\ell \simeq 1$~\cite{Czarnecki:1994pu}. In addition, we consider the $\Lambda_c^+ \to pK^-\pi^+$ decay, which can also be used for spin measurements~\cite{LHCb:2022sck}.

\section{Feasibility of the polarization and spin correlation measurements}

Through MadGraph~\cite{Alwall:2014hca} and Pythia8~\cite{Sjostrand:2014zea} simulations, we estimated the expected statistical uncertainties for measurements in $pp\to b \bar b$ and $pp \to c \bar c$ samples.

\subsection{\texorpdfstring{$pp\to b \bar b$}{pp -> bb} samples}

For this process, we rely on the inclusive semileptonic decay of $\Lambda_b$ with a muon, $\Lambda_b\to X_c\mu^- \bar{\nu}_\mu$.
We impose cuts based on triggers that exist or are planned at ATLAS~\cite{ATL-PHYS-PUB-2019-005}, with the double muon triggers requiring $p_T>15$~GeV for Run~2 and $p_T>10$~GeV for the HL-LHC without isolation requirements. The single muon triggers have $p_T>27$~GeV for Run~2 and $p_T>20$~GeV for the HL-LHC with an efficiency loss of about 50\% due to isolation cuts. Additionally, we apply an $|\eta|<2.4$ cut for Run~2 and $|\eta|<2.5$ for the HL-LHC analyses. To suppress backgrounds, we demand at least one of the jets to be $b$ tagged, and require each muon to carry at least $20\%$ of the jet momentum.

We also found that CMS parked data~\cite{Bainbridge:2020pgi,CMS:2024zhe}, which is data collected during Run~2 that did not pass the standard triggers but was saved for later analysis, provide an excellent dataset for measuring polarization and spin correlations in $b\bar b$. In this data, a single displaced-muon trigger with evolving $p_T$ cuts ($7$--$12$~GeV) and $|\eta| < 1.5$ was applied.

We consider three types of selections (similar to Ref.~\cite{Galanti:2015pqa}):
\begin{itemize}
\item \emph{Inclusive Selection}. This selection, which does not apply any conditions to the particles produced along with the muon, will have high signal efficiency, but also unsuppressed backgrounds from semileptonic decays of $b$ mesons.
\item \emph{Semi-inclusive Selection}, where the $X_c$ is required to contain a $\Lambda \to p\pi^-$ decay, to reduce backgrounds from the semileptonic $b$-meson decays.
\item \emph{Exclusive Selection}, where the $X_c$ is required to contain a fully reconstructible $\Lambda_c^+$ decay (i.e., with charged products only), to reduce backgrounds from the semileptonic meson decays and facilitate the reconstruction of the $\Lambda_b$ decay kinematics for the polarization and spin correlation measurements.
\end{itemize}

In Table~\ref{tab:conclusion}, we show the channels where polarization and spin correlations are measurable. Crossed check marks indicate borderline statistical significance, and parentheses indicate cases with low sample purity (below 10\%), where systematic uncertainties are expected to be significant.  

\begin{table}
\centering
\begin{tabular}{c|c|cc|c} \hline\hline
    &\multicolumn{4}{c}{Polarization} \\\hline
    \multirow{2}{*}{Quark} & \multirow{2}{*}{Channel} & \multicolumn{2}{c|}{Run 2} & \multirow{2}{*}{HL-LHC} \\
    & & standard & parked & \\\hline
    \multirow{3}{*}{$c$}
        & hadronic        & (\halfcheckmark) & & \checkmark\\
	& semileptonic    & \checkmark       & & \checkmark\\
	& mixed           & (\checkmark)     & & \checkmark\\\hline
    \multirow{3}{*}{$b$}
         & inclusive      & (\checkmark) & (\checkmark) & (\checkmark) \\
         & semi-inclusive & \checkmark   & \checkmark   & \checkmark \\
         & exclusive      & \checkmark   & \checkmark   & \checkmark
    \\\hline\hline
\end{tabular}
\vskip 4mm
\centering
\begin{tabular}{c|c|cc|c} \hline\hline
    &\multicolumn{4}{c}{Spin Correlations}\\\hline
    \multirow{2}{*}{Quark} & \multirow{2}{*}{Channel} & \multicolumn{2}{c|}{Run 2} & \multirow{2}{*}{HL-LHC} \\
    & & standard & parked & \\\hline
    \multirow{3}{*}{$c$}
        & hadronic     & & & \\
	& semileptonic & & & \checkmark \\
	& mixed        & & & \checkmark \\\hline
    \multirow{6}{*}{$b$}
	 & inclusive/inclusive           & (\halfcheckmark) & (\checkmark) & (\checkmark) \\
	 & semi-inclusive/semi-inclusive & \halfcheckmark   & \checkmark   & \checkmark \\
	 & exclusive/exclusive           & \halfcheckmark   & \checkmark   & \checkmark \\
	 & inclusive/exclusive           & (\halfcheckmark) & (\checkmark) & (\checkmark) \\
	 & inclusive/semi-inclusive      & (\halfcheckmark) & (\checkmark) & (\checkmark) \\
	 & exclusive/semi-inclusive      & \halfcheckmark   & \checkmark   & \checkmark \\
    \hline\hline
\end{tabular}
\caption{The prospects in Run~2 and HL-LHC datasets for the different analysis channels examined. The top table is for polarization and the bottom is for spin correlations. Check marks indicate that a measurement is expected to be possible, and crossed check marks indicate borderline cases. Parentheses around a check mark indicate low sample purity, under 10\%. The ``parked'' column refers to the CMS parked $b$-hadron dataset~\cite{Bainbridge:2020pgi,CMS:2024zhe}.}
\label{tab:conclusion}
\end{table}

Spin correlations can be used to infer entanglement in the $b\bar b$ final state~\cite{Afik:2024uif} via
\begin{equation}
    \Delta\equiv \frac{-c_{nn}+|c_{kk}+c_{rr}|-1}{2} \,,
\end{equation}
where $\Delta>0$ is a sufficient condition for entanglement, and $k$, $n$, $r$ are a set of orthogonal axes defined in Ref.~\cite{Afik:2024uif}. 
In addition, highly entangled states can exhibit Bell nonlocality. A simple sufficient condition for this is $\mathcal{V}>0$~\cite{Afik:2024uif}, where
\begin{equation}
    \mathcal{V}\equiv c^2_{kk}+c^2_{rr}-1\,.
\end{equation}
We and collaborators have found that entanglement measurements are possible with Run~2 data, and that Bell nonlocality measurements will become possible at the HL-LHC~\cite{Afik:2024uif}. 

\subsection{\texorpdfstring{$pp\to c\bar c$}{pp -> cc} samples}

For $c\bar c$ polarization and spin correlation measurements, we considered three possible analysis channels in terms of the $\Lambda_c^+$ decays: the \emph{hadronic channel} where $\Lambda_c^+ \to pK^-\pi^+$, the \emph{semileptonic channel} where $\Lambda_c^+ \to \Lambda(\to p \pi^-)\mu^+\nu_\mu$, and the \emph{mixed channel} with the hadronic decay in one jet and the semileptonic decay in the other. The semileptonic and mixed channels result in the smallest statistical uncertainties.\footnote{While the branching ratio of the semileptonic decay chain ($3.5\%$ for $\Lambda_c^+\to\Lambda\mu^+\nu_\mu$ and $64\%$ for $\Lambda\to p\pi^-$~\cite{Workman:2022ynf}) is smaller than that of the hadronic decay ($6.3\%$~\cite{Workman:2022ynf}), the muon triggers have low $p_T$ thresholds, as mentioned for the $b\bar b$ case above, while the hadronic decay requires the use of jet triggers with $p_T > 460$~GeV for Run~2 and $400$~GeV for the HL-LHC.}
The cases in which measurements can be done are shown in Table~\ref{tab:conclusion}.

\section{Conclusions}

We show that polarization is measurable with the available Run~2 data for both $c$ and $b$ quarks. Spin correlations are measurable for $b$ quarks with the CMS Run~2 parked data, and will become measurable for $c$ quarks at the HL-LHC. 
These measurements can be used to quantify the polarization retention factors, which in turn provide information on the QCD physics of polarization transfer in fragmentation. They can also be used to measure entanglement and Bell nonlocality, and may even serve as probes for beyond-the-Standard-Model physics, similar to measurements in $t\bar t$.

\bibliographystyle{JHEP}
\bibliography{mybib.bib}

\end{document}